\documentclass{talk}

\begin{document}

\title{Experimental Studies of Diffractive Processes at HERA}

\author{A.A.~Savin \\ on behalf of the H1 and ZEUS collaborations}

\address{
 University of Wisconsin,1150 University Ave.,Madison WI 53706-1390,
 USA\\ 
E-mail: savin@mail.desy.de}


\maketitle

\abstracts{
Diffractive processes in photon-proton interactions at HERA
offer the opportunity to improve the understanding of the
transition
between the soft, non-perturbative regime in hadronic interactions
at $Q^2 = 0$ and
the perturbative region at high $Q^2$. Recent experimental results from HERA on
inclusive diffractive scattering, exclusive vector meson
production and the properties of the hadronic final
state in diffraction are reviewed. The results are discussed in
the context of current theoretical models.
}

\section{Introduction}

\begin{figure}[b]
\begin{center}
\epsfxsize=17pc 
\epsfbox{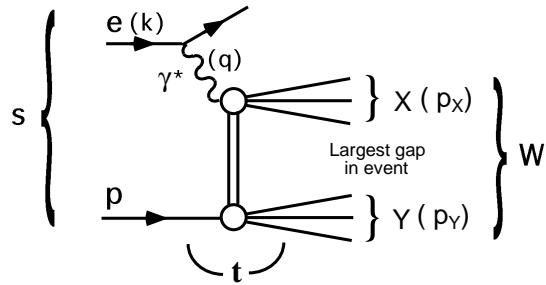} 
\caption{A generic diffractive process at HERA. \label{fig:fd}}
\end{center}
\end{figure}

One of the most important results from the $ep$ collider HERA is the observation
that about 10$\%$ of deep inelastic scattering (DIS) events exhibit a large
rapidity gap between the direction of the proton beam and that of the first
significant deposition in the detector,
thus showing a behaviour typical for diffractive interactions\cite{pl:b315:481,%
np:b429:477}. These events
could be interpreted in terms of the exchange of a colour-singlet object known
as the Pomeron ($\pom$), that can be described in QCD-inspired models as an object
whose partonic composition is dominated by gluons. Alternatively, the
diffractive process can be described by the dissociation of the virtual photon
into a
$q \bar q$ or $q \bar q g$ final state that interacts with the proton by
exchange of a gluon ladder ~\cite{theory}.

Figure ~\ref{fig:fd} illustrates the generic diffractive process at HERA of the type
$e p \rightarrow e X Y $. The positron
couples to a
virtual photon $\gamma ^{*} (q)$ which interacts with the proton ($P$).
Together with the usual kinematic variables like $Q^2, y, W$ and $x$,
one uses $t$, for the 4-momentum transferred at the proton vertex,
$M_X$ and $M_Y$ for the invariant masses of the 
photon and proton dissociative systems, respectively.
$x_{\pom} = \frac {q \cdot ( P - P_{Y} )} {q \cdot P } $ is the fraction of proton
momentum carried by the $\pom$ and $\beta = \frac {x} {x_{\pom}} $ is 
the fraction of the $\pom$ momentum carried by parton 
coupling to $\gamma ^{*}$.

The main  HERA results can be divided into two parts: inclusive
diffraction, such as $F_2^{D(3)}$, jets and hadronic final state studies and 
exclusive diffraction, mainly the vector meson measurements.

\section{Results}

\subsection{Inclusive diffraction}

A high precision inclusive measurement of
$F_2^{D(3)} (\beta, Q^2, x_{\pom})$ was performed recently by H1 in the
kinematic range $6.5 < Q^2 < 120$ GeV$^2$, $0.01 < \beta < 0.9 $ and
$10^{-4} < x_{\pom} < 0.05 $. The $x_{\pom}$ dependence of the data
was interpreted in terms of a measurement of the effective pomeron
intercept $\alpha_{\pom} (0)$ that was found to be

$\alpha_{\pom} (0) = 1.173 \pm 0.018 (stat.) \pm 0.017 (syst.) ^{+0.063}
_{-0.035} (model) $. 

\begin{figure}[htb]
\begin{center}
\epsfxsize=17pc 
\epsfbox{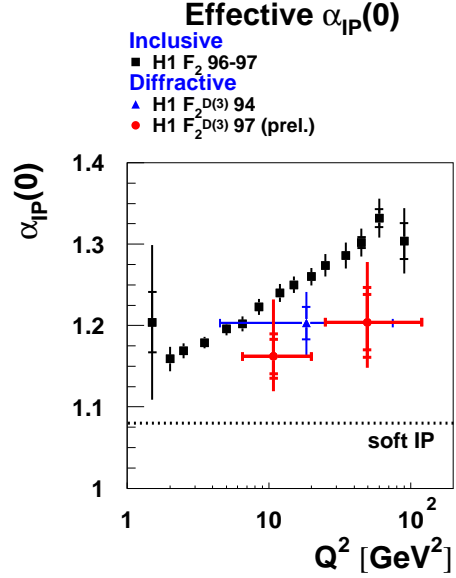} 
\caption{The effective value of $\alpha_{\pom}(0)$ as a function
of $Q^2$. The squares correspond to $\alpha_{\pom}(0) = 1 + \lambda$
extracted from a fit $F_2 = cx^{-\lambda(Q^2)}$ to inclusive
$F_2(x,Q^2)$ ~\protect\cite{h1incl} data for $x<0.01$. The filled
circles are the values of $\alpha_{\pom}(0)$ as obtained from
the phenomenological Regge fit to the $F^D_2$ data as described
in the text. The triangle is the value of $\alpha_{\pom}(0)$ 
which was already measured by H1 using 94' data ~\protect\cite{f2d3:old}
\label{fig:h1} }
\end{center}
\end{figure}

Two further fits were performed in order to investigate
whether $\alpha_{\pom}(0)$ has any dependence on $Q^2$, the results are
shown in Fig.~\ref{fig:h1}. 
The values are significantly higher than the value 
 $\alpha_{\pom} (0) = 1.08 $ for the $soft~ pomeron$ ~\cite{pomeron},
 but apparently lower than for inclusive $F_2(x, Q^2)$
 measurements.

\begin{figure}[t]
\epsfxsize=30pc 
\epsfbox{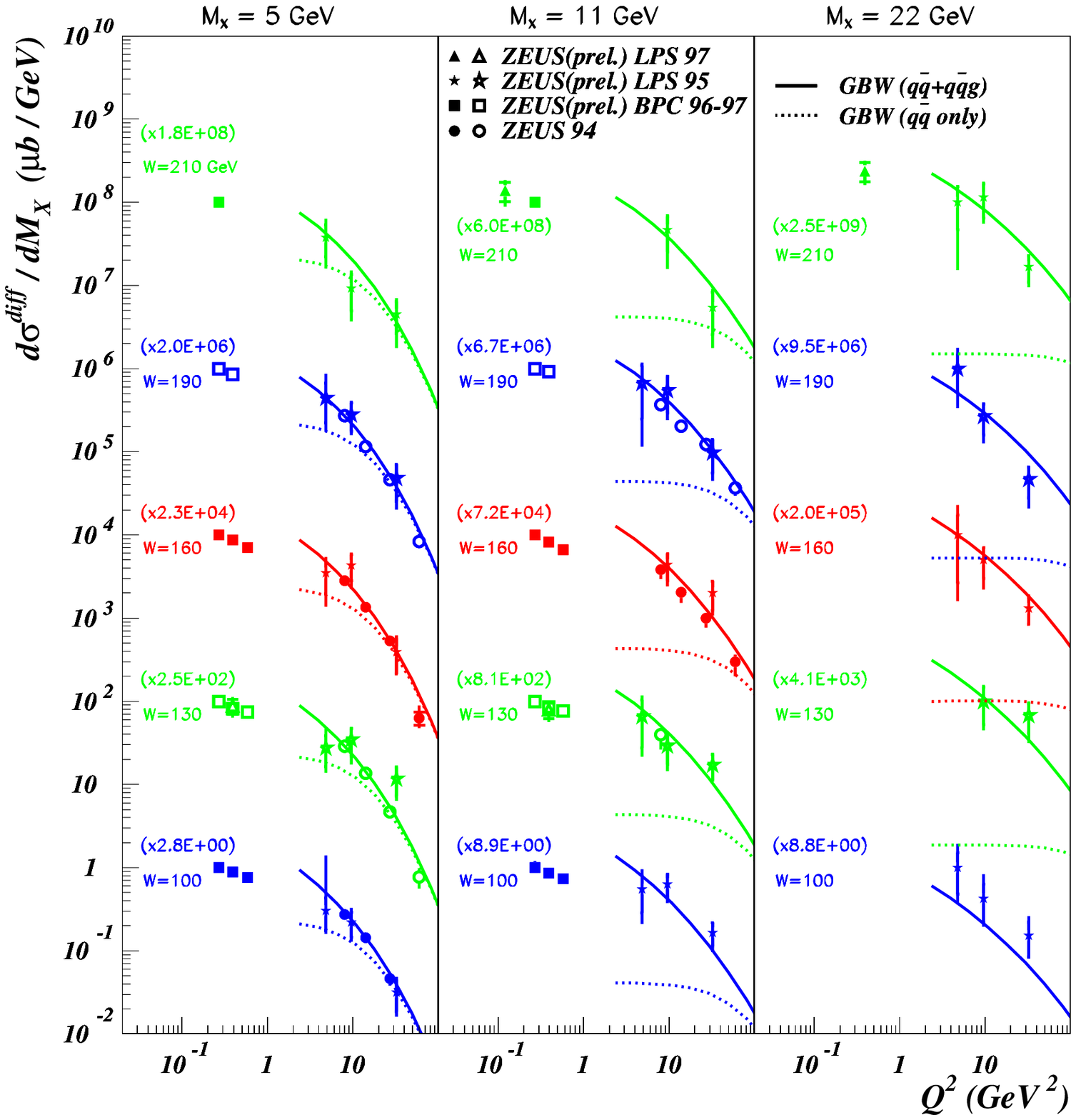} 
\caption{The diffractive cross sections for 
different $W$ and $M_X$ values as a function of $Q^2$ 
The continuous curves indicate the predictions of the model~\protect\cite{GBW} 
for $q \bar q + q \bar q g $
fluctuations, the dashed ones are the predictions 
if only $ q \bar q$ fluctuations
are considered. \label{fig:dsdmx} }
\end{figure}

 At fixed $x_{\pom}$, a flat $\beta$ and a rising $Q^2$ dependencies were 
 observed. This structure was well described by a fit based on DGLAP evolution
 with diffractive parton distributions of the proton heavly dominated
 by a large gluon density. This observation is in good agreement with the 
 previous $F_2^{D(3)}$ measurements ~\cite{f2d3:old}.

Recently ZEUS updated the diffractive cross section measurement at very 
low $Q^2$.
The cross sections for different $W$ and $M_X$ values as a function of $Q^2$
are compared with the predictions of the saturation dipole-model ~\cite{GBW}
 in Fig.~\ref{fig:dsdmx}. 
The continuous curves indicate the model predictions for $q \bar q + q \bar q g $
fluctuations, the dashed ones are the predictions if only $ q \bar q$ 
fluctuations are considered. It is obvious that the 
$ q \bar q$ configuration alone does not describe the data.

\begin{figure}[t]
\epsfxsize=13pc 
\epsfbox{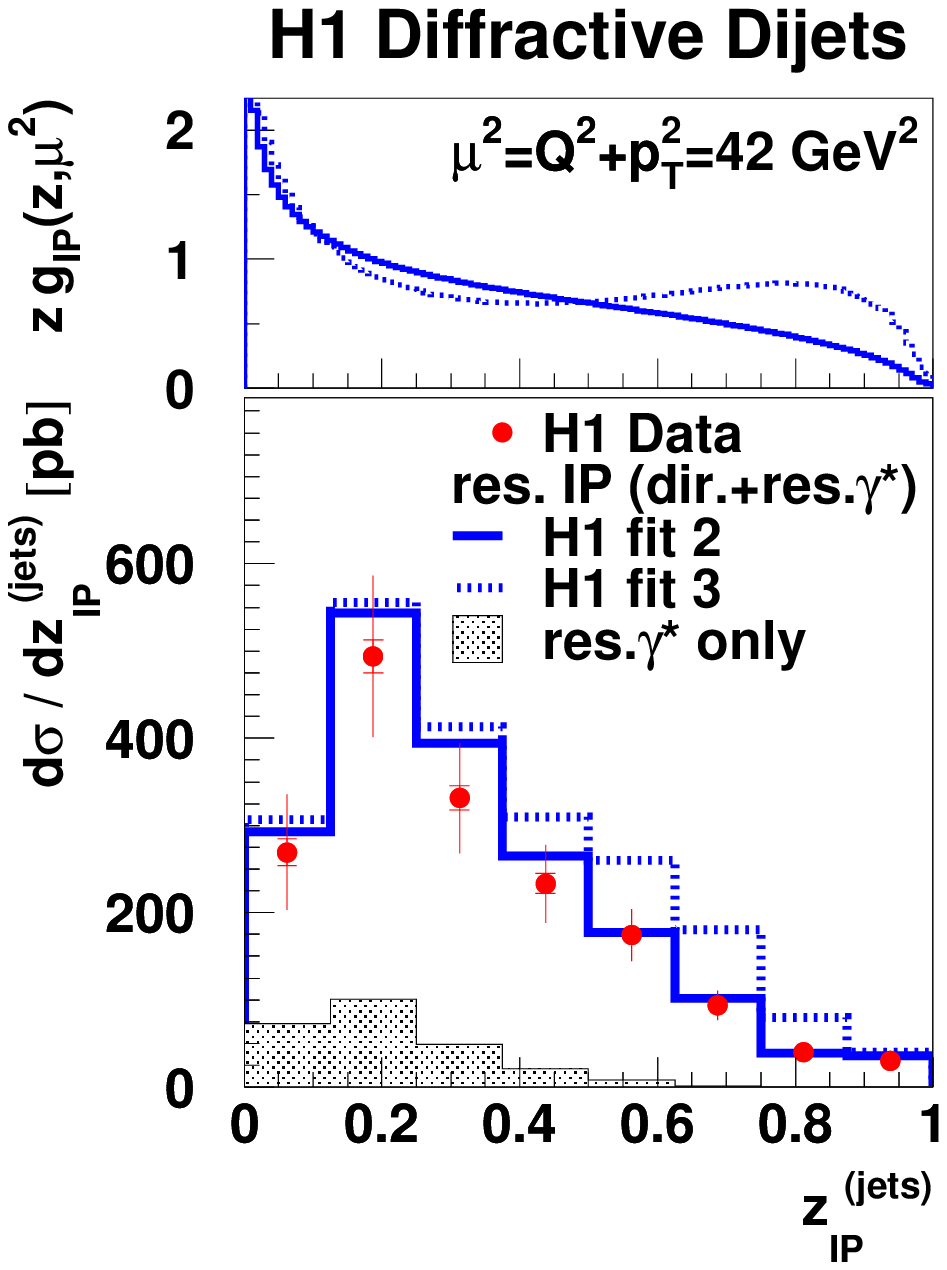} 
\epsfxsize=18pc 
\epsfbox{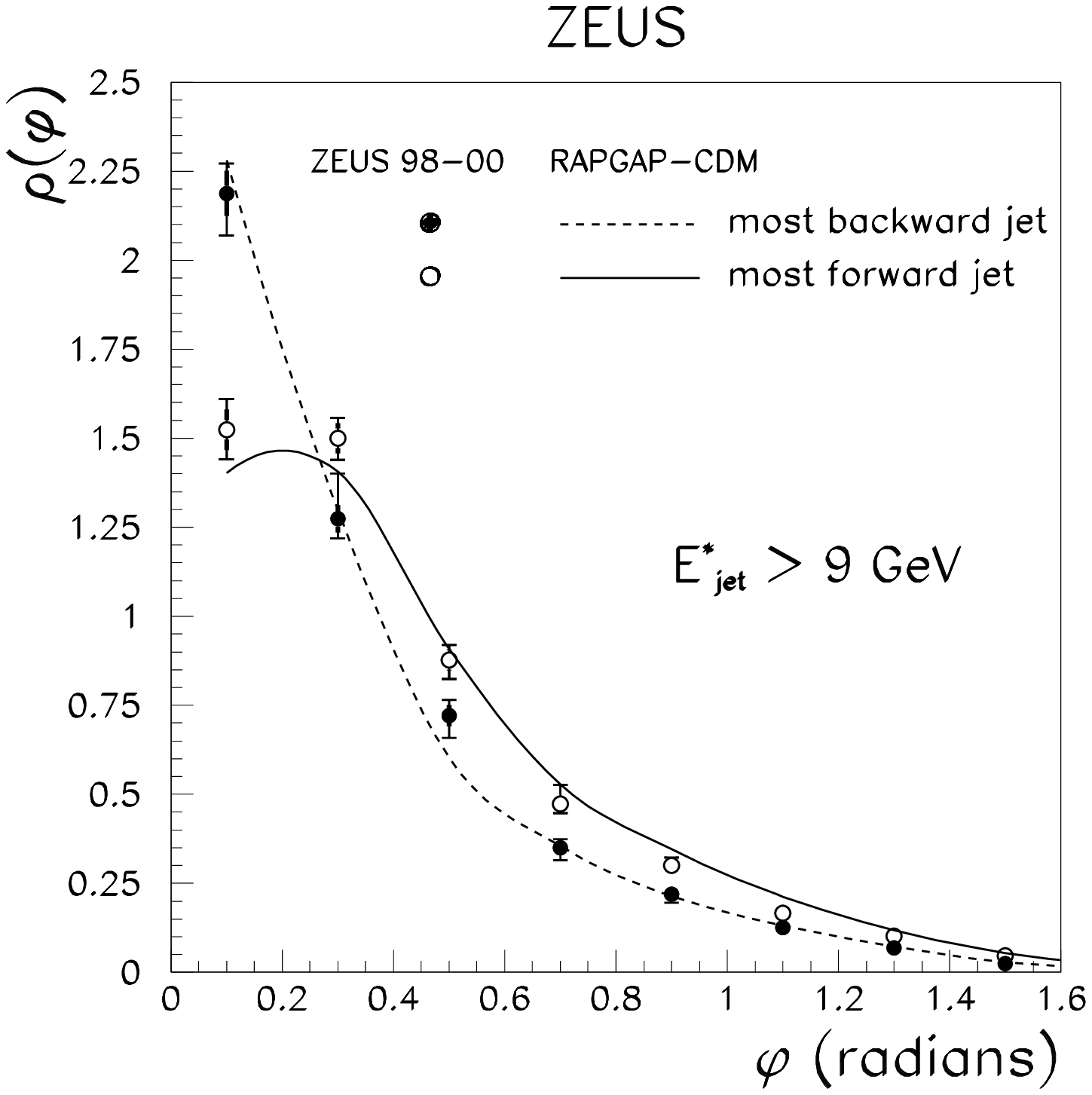} 
\caption{a) The diffractive dijet cross section as a function of
$z^{(jets)}_{\pom}$. The same data are compared to predictions
of resolved pomeron models. The fits are based on the H1 leading
order QCD fits to $F_2^{D(3)}$~\protect\cite{f2d3:old}.
b) The differential jet shape, $\rho(\varphi)$, for the most-forward and 
most-backward jets in three-jet events, where the Pomeron defines the forward
direction.  \label{fig:jets}}
\end{figure}

 A measurement of dijet cross sections yields direct constrain on the
 diffractive gluon distribution and was used to investigate the QCD and Regge
 factorization properties of diffractive DIS. 
 
 In Fig.~\ref{fig:jets}a) the diffractive dijet cross section as a function of
 $z^{(jets)}_{\pom}$ is shown, where $z^{(jets)}_{\pom} = \frac
 {Q^2 + M^2_{12}} {Q^2 + M^2_X} $ and $M^2_{12}$ is the  squared dijet
 invariant mass. In loose terms, the $z^{(jets)}_{\pom}$  observable measures
 the fraction of the total hadronic final state energy of the $X$ system that is
 contained in the two jets. Diffractively scattered $ q \bar q$ photon
 fluctuations satisfy $ z_{\pom} = 1$ at the parton level and can be
 smeared to $z^{(jets)}_{\pom}$  values as low as 0.6 because of fragmentation and
 jet resolution effects. Even taking this smearing into account, the $z^{(jets)}_{\pom}$ 
 distribution obviously implies the dominance of $ q \bar q g$ over $ q \bar q$ scattering
 in the proton rest frame picture. The distribution can also be very well fitted
 using  the parametrization based on the H1 leading order QCD fits to
 $F_2^{D(3)}$.
 
 Making a fit to the shape of the $x_{\pom}$ dependence
 of the cross section yields a value of 
 $\alpha_{\pom} (0) = 1.17 \pm 0.03 (stat.) \pm 0.06 (syst.) ^{+0.03}
_{-0.07} (model) $ for the Pomeron intercept, very close to the value
 measured using the $F_2^{D(3)}$ data.
 
 The internal structure of the jets was studied in diffractive three-jet  
 production in terms of differential jet shape ~\cite{3jets}, defined as the average of the fraction
 of the jet energy which lies inside an annulus of inner angular distance
 $\varphi - \delta \varphi$ and outer angular distance $\varphi + \delta \varphi$
 from the jet axis. This is an interesting observable because 
 gluon jets are known to be broader than quark jets. The differential jet shapes,
 $\rho (\varphi)$, were measured in the $\gamma ^* \pom$ center-of-mass frame, 
 for the most-forward and most-backward
 jet in three-jet events, see Fig.~\ref{fig:jets}b),
 where the forward region is defined by the
 $\pom$ direction. These measurements are described 
 by models in which a
 gluon populates the Pomeron hemisphere and a quark is found in the
 photon direction, in good agreement with our knowlege about
 diffractive dynamics from the other inclusive measurements.

\begin{figure}[t]
\epsfxsize=14pc 
\epsfbox{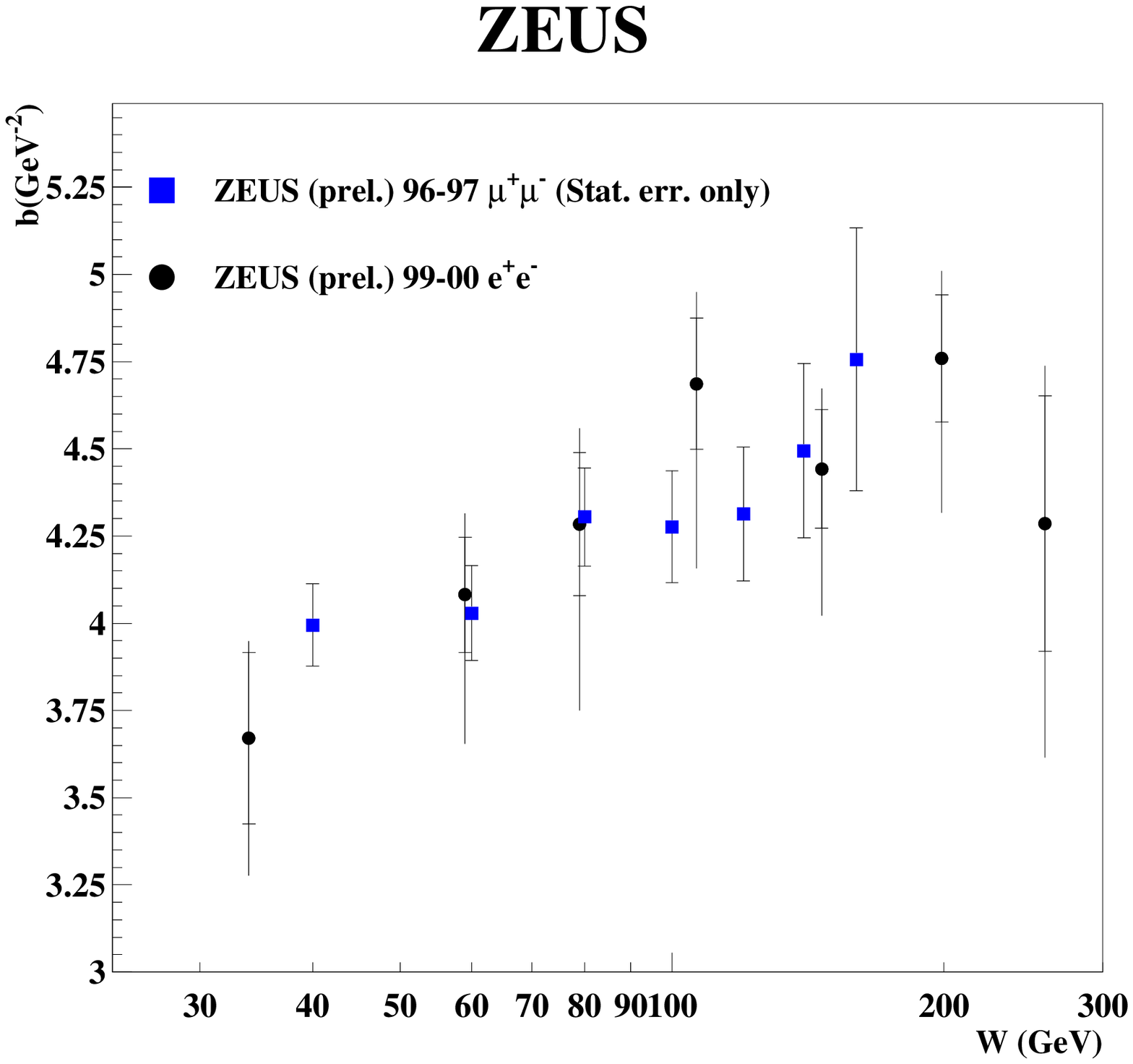} 
\epsfxsize=14pc 
\epsfbox{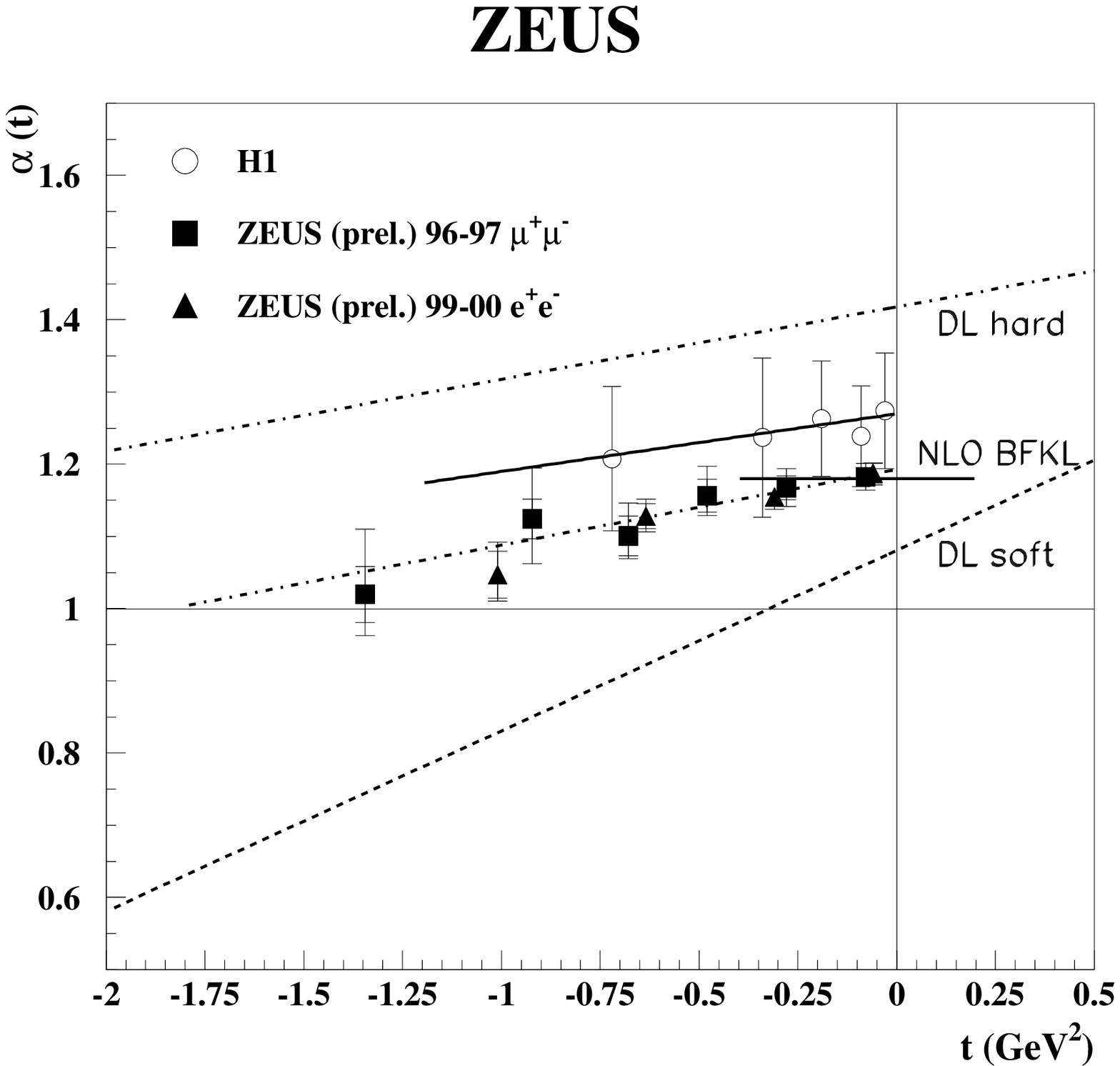} 
\caption{a) The $b$ slope for the $J/\psi$ photoproduction cross section versus
$W$ ; 
b) The Pomeron trajectory as a function of $t$ from the H1~\protect\cite{h1jpsi}
and ZEUS experiments.
The inner error bars indicate the  statistical uncertainties errors, 
the outer bars are the statistical
and systematic uncertainties added in quadrature.
The result of straight line fits to the H1 and ZEUS data are shown.
Also shown are the soft (dashed line) and the hard 
(dot-dashed line) DL Pomeron trajectories~\protect\cite{DLhard}
 and a prediction (shown by the solid line) for the 
Pomeron intercept based on a NLO BFKL calculation~\protect\cite{brodsk}.
 \label{fig:jpsi}}
\end{figure}

To summarize the $Inclusive$ section, one can conclude that the data
are broadly consistent with models in which the diffractive 
hadronic final-state is dominated by a $q \bar q g$ system with the gluon
preferentally emitted in the Pomeron direction. Both the resolved Pomeron 
models with a pomeron, dominated by gluon, as well as the models where the
virtual photon dissociates to a $ q \bar q g $ system, which interacts with 
the proton via the exchange of a gluon ladder, provide a reasonable 
description of the measured distributions.

\subsection{Exclusive diffraction}

The exclusive, diffractive production of vector mesons, $ep \rightarrow eVp$,
where $V=( \rho ^0;~ \omega;~ \phi;~ J/\psi;~ \psi ';~ \Upsilon) $ is 
measured at HERA in the region of $0 < Q^2 < 100$ GeV$^2$, 
$20 < W_{\gamma P} < 290$ GeV, $0 < |t| < 1.5$ GeV$^2$ and up to $|t| = 20$
~GeV$^2$ for proton-dissociative events.

Photoproduction of $J/\psi$ mesons ($Q^2 \approx 0 $), 
contrary to the photoproduction of the light vector mesons, shows the signature
of the perturbative regimes such as a steep rise of the total cross section with
$W$, and can be described by perturbative QCD, since the mass of the charm
quark provides a hard scale, similar to the $Q^2$ in the DIS case. At the same 
time, 
QCD models predict no variation of the t-dependence of the cross section with
$W$. 

Fitting the differential cross section as  
$\frac {d \sigma } {dt} \propto e^{-bt} $ the values of the 
slope $b$ in different bins of $W$ are obtained, 
see Fig.~\ref{fig:jpsi}a). 
The data show the rise of the $b$ slope with $W$, so called
$shrinkage$, contrary to the pQCD expectation.
  
The photoproduction of $J/\psi$ mesons can also be described within
the framework of Regge phenomenology~\cite{collins}. In this approach,
$d \sigma / dt$ can be expressed at high energies as
$ \frac {d \sigma} {dt} \propto e^{bt} \cdot W^{2 \cdot [2 \alpha_{\pom}-2]}$,
where $\alpha_{\pom} (t) =
\alpha_{\pom} (0) + \alpha_{\pom}' t $ is the Pomeron trajectory.
The $W$ dependence of $b$ can be used then to estimate $\alpha_{\pom}'$,
since $ b = b_0 + 2 \alpha_{\pom}' ln(W/W_0)^2 $. A fit to the data
yields the
$\alpha_{\pom} ' = 0.122 \pm 0.033 (stat.) \pm ^{+0.018}_{-0.032} (syst.) $
 GeV$^{-2}$, that is much smaller than 0.25 of the  $soft~ Pomeron$.
  
  The Pomeron trajectory is also determined by measuring the variation of the
  energy dependence of the cross section at fixed $t$. 
  The resulting values
  of $\alpha_{\pom} (t)$ are presented in Fig.~\ref{fig:jpsi}b) as a function of $t$ and are
  fitted to the linear form yielding (the recent ZEUS data)

  $\alpha_{\pom} (0) = 1.201 \pm 0.013 (stat.) \pm ^{+0.003}_{-0.011} (syst.) $ 
  and
  
  $\alpha_{\pom} ' = 0.126 \pm 0.029 (stat.) \pm ^{+0.015}_{-0.028} (syst.) $
 GeV$^{-2}$.

  Analysis of the HERA vector meson cross-section ratios results shows 
  that $Q^2, M_{VM}^2$ and $t$
  can play a role of the hard scale in diffractive scattering 
  ~\cite{h1phi,zeust}. 
  The cross-section dependence is similar for  $Q^2$ and $M_{VM}^2$
  and suggests using a combined scale ($Q^2 + M^2_{VM}$), but the 
  $t$ dependence 
  is different. Existing pQCD models provide reasonable 
  agreement with experimental data.

To summarize the $Exclusive$ section, one can conclude that the data
are consistent with the inclusive diffractive measurements, showing
a larger value of the Pomeron intercept and a smaller value 
of the $\alpha_{\pom} '$ in the presence of the hard scale.


\begin{thebibliography}{99}
\bibitem{pl:b315:481}
{ZEUS \coll, M.~Derrick \etal,%
 ~Phys.\ Lett.{} {\bf B~315},~481~(1993)}
\bibitem{np:b429:477}
{H1 \coll, T.~Ahmed \etal,%
 ~Nucl.\ Phys.{} {\bf B~429},~477~(1994)}
\bibitem{theory}
{See e.g. {\it Proc. of the Workshop on Future Physics at HERA},%
~G.~Ingelman, A.~DeRoeck and R.~Klanner (eds.), Vol.2, DESY, Hamburg, (1996)}%
\bibitem{h1incl}
{H1 \coll, C.~Adloff \etal,%
~Eur.\ Phys.\ J.{} {\bf C~21},~33~(2001)}
\bibitem{pomeron}
{A.~Donnachie and P.V.~Landshoff,%
~Phys.\ Lett.{} {\bf B~296},~227~(1992)}
\bibitem{f2d3:old}
{H1 \coll, C.~Adloff \etal,%
~Z.\ Phys.{} {\bf C~76},~613~(1997)}
\bibitem{GBW}
{K.~Golec-Biernat and M.~W{\"u}sthoff,%
~Phys.\ Rev.{} {\bf D~60},~114023~(1999)}
\bibitem{3jets}
{ZEUS \coll, S.~Chekanov \etal,%
 ~Phys.\ Lett.{} {\bf B~516},~273~(2001)}
\bibitem{collins}
{P.D.B.~Collins, {\it An introduction to Regge Theory and Hogh Energy Physics},%
~Cambrige University Press, 1997 }
\bibitem{h1jpsi}
{H1 \coll, C.~Adloff \etal,%
~Phys.\ Lett.{} {\bf B~483},~23~(2000)}
\bibitem{DLhard}
{A.~Donnachie and P.V.~Landshoff,%
~Phys.\ Lett.{} {\bf B~470},~243~(1999) and references therein.}
\bibitem{brodsk}
{S.J.~Brodsky \etal,%
~Sov.\ Phys.\ JETP{} {\bf 70},~155~(1999)}
\bibitem{h1phi}
{H1 \coll, C.~Adloff \etal,%
~Phys.\ Lett.{} {\bf B-483},~360~(2000)}
\bibitem{zeust}
{ZEUS \coll, S.~Chekanov \etal,%
~Abstr.556, {\it subm. to the EPS 2001, Budapest}%
~2001}
\end{thebibliography}
\end{document}